\documentclass[
  journal=pasa,
  manuscript=research-paper, 
  year=2020,
  volume=37,
]{cup-journal}

\usepackage{microtype,siunitx,booktabs}

\usepackage{graphicx}
\usepackage{amsmath}	
\usepackage{amssymb}	
\usepackage{multicol}       
\usepackage{bm}		
\usepackage{pdflscape}	
\usepackage{xurl}
\usepackage{hyperref} 
\usepackage[T1]{fontenc}
\usepackage{ae,aecompl}
\usepackage{xcolor} 
\usepackage{booktabs}
\usepackage{pifont}
\usepackage{nicematrix}
\usepackage{longtable}
\usepackage{tabularx}
\usepackage{booktabs}
\usepackage{caption}

\DeclareMathOperator*{\argmax}{arg\,max}

\title[EMUSE]{EMUSE: Evolutionary Map of the Universe Search Engine}

\author{Nikhel Gupta$^{1}$} 
\author{Zeeshan Hayder$^{2}$}
\author{Minh Huynh$^{1,5}$}
\author{Ray P. Norris$^{3,4}$}
\author{Lars Petersson$^{2}$}
\author{Andrew M. Hopkins$^{6}$}
\author{Simone Riggi$^{7}$}
\author{B\"arbel S. Koribalski$^{4,3}$}
\author{Miroslav D. Filipovi\'c$^{3}$}
\email[Nikhel Gupta]{Nikhel.Gupta@csiro.au}
\affiliation{
$^1$ Australia Telescope National Facility, CSIRO, Space \& Astronomy, PO Box 1130, Bentley WA 6102, Australia \\
$^2$ CSIRO Data61, Black Mountain ACT 2601, Australia \\
$^3$ Western Sydney University, Locked Bag 1797, Penrith, NSW 2751, Australia \\
$^4$ Australia Telescope National Facility, CSIRO Space \& Astronomy, P.O. Box 76, Epping, NSW 1710, Australia \\
$^5$ International Centre for Radio Astronomy Research (ICRAR), M468, The University of Western Australia, 35 Stirling Highway, Crawley, WA 6009, Australia \\ 
$^6$ School of Mathematical and Physical Sciences, 12 Wally’s Walk, Macquarie University, NSW 2109, Australia \\
$^7$ INAF-Osservatorio Astrofisico di Catania, Via Santa Sofia 78, 95123 Catania, Italy \\
}

\received {dd Mmm YYYY}
\revised  {dd Mmm YYYY}
\accepted {dd Mmm YYYY}
\published{dd Mmm YYYY}


\usepackage{amsmath,amsfonts,amssymb}
\usepackage{color}
\usepackage{mathrsfs}
\usepackage{tabularx}
\usepackage{floatrow}
\usepackage{float}

\definecolor{ored}{rgb}{1.00,0.27,0.00}
\definecolor{mygreen}{rgb}{0.2,0.7,0.2}

\definecolor{Gray}{gray}{0.5}
\definecolor{LightCyan}{rgb}{0.88,1,1}

\def \BE{\begin{equation}}
\def \EE{\end{equation}}	
\def \BC{\begin{center}}
\def \EC{\end{center}}
\def \BEA{\begin{eqnarray}}
\def \EEA{\end{eqnarray}}

\def \SIGMA8{\sigma_{8}}

\usepackage{cellspace}
\begin{document}\sloppy\sloppypar\raggedbottom\frenchspacing

\begin{abstract}
We present EMUSE (Evolutionary Map of the Universe Search Engine), a tool designed for searching specific radio sources within the extensive datasets of the EMU (Evolutionary Map of the Universe) survey, with potential applications to other Big Data challenges in astronomy.
Built on a multimodal approach to radio source classification and retrieval, EMUSE fine-tunes the OpenCLIP model on curated radio galaxy datasets.
Leveraging the power of foundation models, our work integrates visual and textual embeddings to enable efficient and flexible searches within large radio astronomical datasets. 
We fine-tune OpenCLIP using a dataset of 2,900 radio galaxies, encompassing various morphological classes, including FR-I, FR-II, FR-x, R-type, and other rare and peculiar sources. 
The model is optimised using adapter-based fine-tuning, ensuring computational efficiency while capturing the unique characteristics of radio sources. 
The fine-tuned model is then deployed in the EMUSE, allowing for seamless image and text-based queries over the EMU survey dataset. 
Our results demonstrate the model’s effectiveness in retrieving and classifying radio sources, particularly in recognising distinct morphological features. 
However, challenges remain in identifying rare or previously unseen radio sources, highlighting the need for expanded datasets and continuous refinement. 
This study showcases the potential of multimodal machine learning in radio astronomy, paving the way for more scalable and accurate search tools in the field. 
The search engine is accessible at \url{https://askap-emuse.streamlit.app/} and can be used locally by cloning the repository at \url{https://github.com/Nikhel1/EMUSE}.
\end{abstract}

\section{Introduction}
\label{SEC:Intro}
The Evolutionary Map of the Universe \citep[EMU;][]{hopkins25} survey, conducted with the Australian Square Kilometre Array Pathfinder \citep[ASKAP;][]{johnston07ASKAP,DeBoer09,hotan21}, highlights the transformative role of modern radio interferometers in cosmic exploration. 
Over its five-year duration, the survey aims to detect more than 20 million compact and extended radio galaxies, providing an unprecedented dataset that will significantly enhance our understanding of galaxy evolution and the Universe’s history. 
Additionally, such extensive data are expected to unveil new astrophysical phenomena and offer deeper insights into the origins of radio emissions. 
However, achieving these scientific objectives requires moving beyond conventional data mining techniques. 
Instead, innovative approaches are needed to analyse, organise, and classify the vast amounts of radio galaxy data, leveraging multiwavelength observations to unlock the survey’s full potential.

In recent years, machine learning has become a powerful tool for analyzing data from the next generation of radio telescopes \citep[e.g.][]{mostert21, gupta22, walmsley22, segal22, alegre22, gupta23a, lochner23, gupta23b, slijepcevic23, Mohale24, gupta24a, Lastufka24, gupta24b, riggi24, lochner24, mostert24, lao25, gupta25a}. 
These techniques have significantly accelerated both the discovery of new radio morphologies and the detection, classification, and cataloguing of radio sources.
Beyond the approaches employed in these studies, emerging models with multimodal capabilities offer new opportunities to enhance the analysis of Big Data from radio telescopes. 
For instance, foundation models, which are large-scale deep learning architectures pre-trained on diverse datasets, can be adapted for radio astronomy tasks. 
These models, such as Generative Pre-training Transformer \citep[GPT;][]{brown20}, Contrastive Language-Image Pre-training \citep[CLIP;][]{radford21}, and vision-language models like Gemini \citep{ gemini23}, have demonstrated remarkable capabilities in cross-modal understanding and pattern recognition.
By leveraging foundation models, we can further improve the detection, classification, and retrieval of radio sky data. 
Their ability to integrate information from multiple data modalities (e.g., radio, infrared, optical) enables more robust source identification and classification \citep[e.g.][]{jia21, alayrac2022flamingo, radford21, ramesh22, rombach22}. 
Additionally, their adaptability through fine-tuning and zero-shot\footnote{An approach where a model is trained to recognise or classify objects, concepts, or tasks it has never seen during training.} learning \citep[e.g.,][]{bommasani21, yu22, touvron23} allows for more efficient exploration of large-scale surveys, making them valuable tools for future radio astronomy research.

Pre-training multimodal foundation models requires vast image-text datasets and significant computational resources.
The lack of open-source models in this domain further hinders progress.
Recently, \citet{parker24} pre-trained a multimodal model on galaxy data using optical imaging and spectral information, applying it to downstream tasks.
Similarly, \citet{riggi25} pre-trained a small vision language model on radio images and image-caption pairs with a focus on downstream generative tasks.
However, research on multimodal model pretraining suggests that while pretraining strategies influence downstream performance, the primary objective of pre-training should be to develop robust, generalizable features rather than domain-specific ones.
Domain adaptation is generally more effective when achieved through fine-tuning on task-specific datasets \citep[see, e.g.,][]{fayou2024clustering,manzoor2023multimodality}.
Notably, \citet{tanoglidis24} employed GPT-4o and LLaVA-NeXT pre-trained models for zero-shot classification of low-surface-brightness galaxies and artifacts, as well as for morphological galaxy classification. Their findings indicate that, with natural language prompts, these models achieved high classification accuracy (typically above 80\%) without additional fine-tuning.
Thus, leveraging a pre-trained model trained on general real-world data is a promising approach for fine-tuning domain-specific tasks while eliminating pre-training costs.
In a recent work, \citet{cherti23} trained CLIP using the public LAION dataset \citep{schuhmann22laion}, which includes an English image-text subset of 2.32 billion real-world samples, to produce OpenCLIP---a large, publicly available image-text model---using approximately 1,520 NVIDIA A100 GPUs. 
This enables the design of downstream tasks using OpenCLIP as a foundation model pre-trained on a vast image-text dataset.

In this work, we develop a framework to fine-tune the OpenCLIP model on the RadioGalaxyNET dataset \citep[][]{gupta24a} derived from the Evolutionary Map of the Universe first pilot survey \citep[EMU-PS1][]{norris21} using a single H100 GPU. 
We then leverage the fine-tuned model to develop EMUSE\footnote{\url{https://github.com/Nikhel1/EMUSE}} (Evolutionary Map of the Universe Search Engine), an application that performs similarity search on the first-year observations of the EMU main survey \citep[][]{hopkins25}. 
EMUSE enables users to explore data and identify similar radio sources through image or text-based queries, allowing for rapid searches of specific radio source classes. This capability is crucial for building statistically robust samples of well-known categories, such as FR-I and FR-II galaxies, as well as for discovering additional examples of rare and peculiar systems. Such samples are essential for investigating population properties, analysing the distribution of morphological types, and tracing their evolution across cosmic time.
Additionally, EMUSE lays the groundwork to develop advanced tools for rapidly extracting meaningful insights and discovering new phenomena from the Big Data produced by next-generation multiwavelength surveys.

The paper is organised as follows. In Section~\ref{SEC:dataset}, we provide details on the EMU survey, infrared observations and object detection-based EMU catalogues. Section~\ref{SEC:model} is dedicated to the foundation models and our fine-tuning approach. Section~\ref{SEC:emuse} provides comprehensive information about the EMUSE application. Our findings are summarised in Section~\ref{SEC:conclusions}, where we also outline directions for future research.

\section{Data}
\label{SEC:dataset}
This section presents an overview of the EMU survey, infrared observations, and the catalogues generated through object detection used in this study.

\subsection{EMU Observations}
\label{SEC:dataset1}
The Evolutionary Map of the Universe (EMU) \citep[EMU\footnote{\url{https://emu-survey.org/}};][]{hopkins25} is a large-scale radio survey being conducted with the Australian Square Kilometre Array Pathfinder \citep[ASKAP;][]{hotan21} to map the southern sky. ASKAP, located at Inyarrimahnha Ilgari Bundara, MRO, consists of 36 antennas, with most within a 2.3 km diameter and six extending to 6.4 km baselines. 
The survey includes 853 tile footprints from 1,014 observations, with 692 tiles having 10-hour integrations and 161 tiles observed twice for 5-hour integrations. EMU covers declinations from $-11^{\circ}.4$ to the south celestial pole and selected equatorial regions up to $\delta = +7^{\circ}.0$, observing in the 800–1088 MHz band, centred at 944 MHz. 
The RMS noise ranges from $25$ to $55~\mu$Jy/beam, with a $13^{\prime\prime} \times 11^{\prime\prime}$ beamwidth. By 2028, EMU aims to detect up to 20 million radio sources over $2\pi$ sr of the sky.
This study uses data from EMU's first-year observations \citep[see][for details]{gupta25a}, covering 160 tiles (~4,500 square degrees). 
Data collection commenced in late 2022, with validated data arriving between February 2023 and March 2024. 
The dataset, accessed via the CSIRO Data Access Portal (CASDA\footnote{\url{https://research.csiro.au/casda/}}), consists of image tiles and \textit{Selavy}-based catalogues \citep[][]{whiting12} with Scheduling Block IDs (SBID) from 45638 to 59612. 
We use restored images at a uniform $15^{\prime\prime}$ resolution per beam (identified by the "conv" filename suffix in CASDA).
For the 160 tiles in the first-year dataset, this amounts to approximately 3 million detected radio sources. 
Each tile is analysed independently rather than combined into super mosaics, which may lead to duplicate detections in overlapping regions. 

\subsection{Infrared Observations}
\label{SEC:dataset2}
In addition to the EMU observations, we generate corresponding 160 tiles for the AllWISE dataset from the Wide-field Infrared Survey Explorer (WISE) \citep[][]{wright10, cutri13} using the Montage image mosaic software\footnote{Implementation available at: \url{https://github.com/Nikhel1/wise_mosaics}}. 
WISE conducted an all-sky infrared survey across four bands--W1, W2, W3, and W4--at wavelengths of 3.4, 4.6, 12, and 22 $\mu$m, respectively. 
This study focuses on the W1 band from AllWISE, which provides a 5$\sigma$ point source detection limit of 28 $\mu$Jy and an angular resolution of $8.5^{\prime \prime}$.

\begin{figure*}[!ht]
\centering
\includegraphics[width=18cm, scale=0.5]{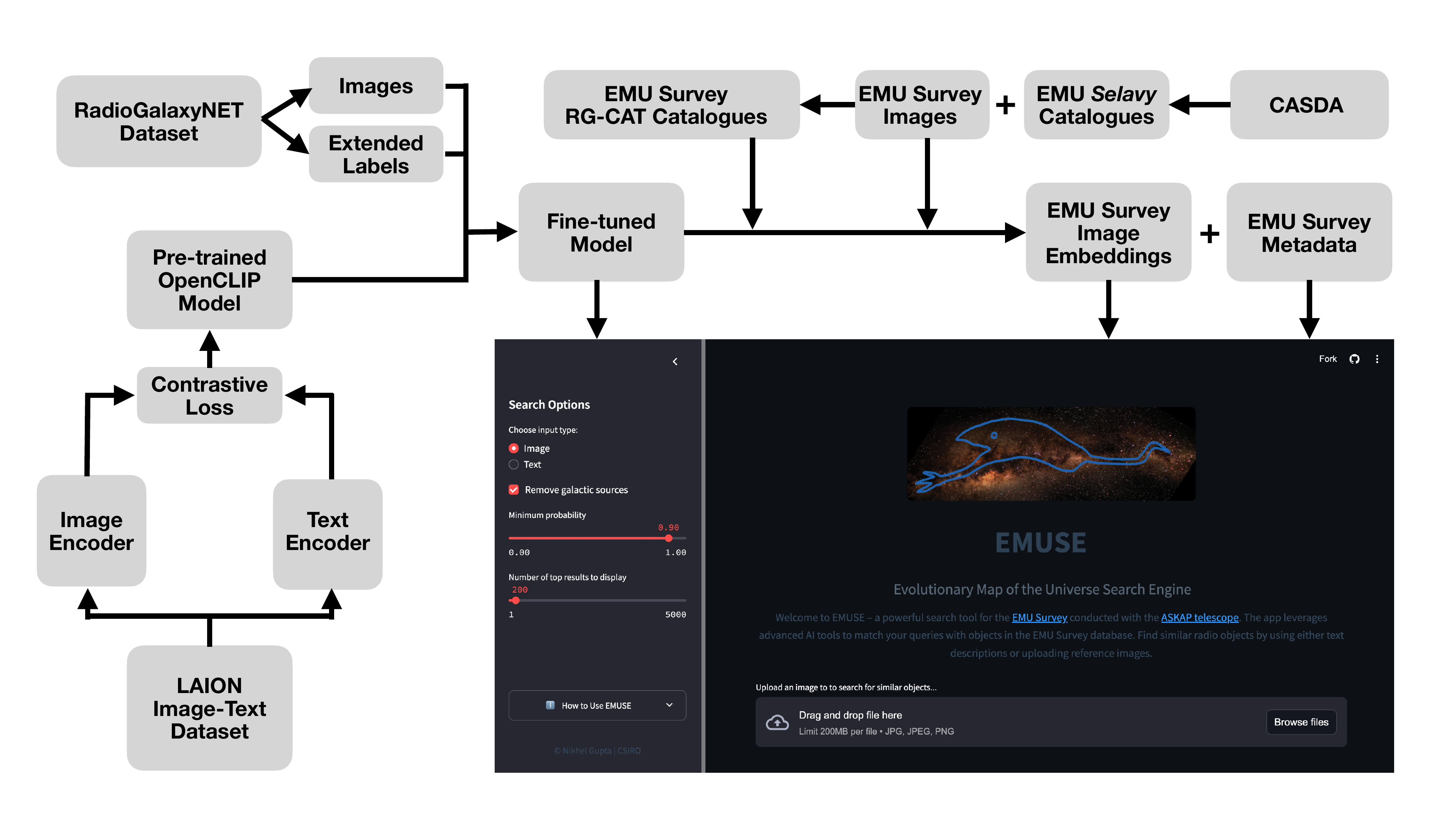}
\caption{Overview of EMUSE (Evolutionary Map of the Universe Search Engine). Starting with the open-source OpenCLIP model, which is pre-trained on approximately 2.3 billion image-text pairs from the LAION dataset, we further fine-tuned it using an image-text dataset of extended radio sources in the EMU-PS1 survey. The fine-tuned model is then used to generate image embeddings of EMU sources based on PNG images from the EMU and AllWISE surveys at the positions of extended radio sources identified in the RG-CAT catalogue. The fine-tuned model, along with the generated image embeddings and catalogue metadata -- which includes sky position, integrated flux, and host galaxy information -- is integrated into the EMUSE application framework to retrieve similar sources. EMUSE facilitates the search of the embedding database and outputs a table of EMU survey radio sources that are similar to a given image or text prompt. The search engine is accessible at \url{https://askap-emuse.streamlit.app/} and can be used locally by cloning \url{https://github.com/Nikhel1/EMUSE}.} 
\label{FIG:EMUSE}
\end{figure*}

\subsection{Catalogues from RG-CAT Pipeline}
\label{SEC:dataset3}
We use the RG-CAT catalogue construction pipeline \citep{gupta24b}, which integrates the Gal-DINO\footnote{\url{https://github.com/Nikhel1/Gal-DINO}} object detection framework \citep[][]{gupta24a} to catalogue radio sources systematically. 
Gal-DINO is designed to detect radio galaxies and identify their probable infrared hosts. 
It is trained on 5,000 radio galaxies, including 2,800 from the RadioGalaxyNET dataset \citep[][]{gupta24a}, spanning FR-I, FR-II, FR-x, and R-type classifications based on peak separation and total extent \citep{fanaroff74}. 
FR-I galaxies have a peak-to-extent ratio below 0.45, FR-II above 0.55, FR-x between 0.45 and 0.55, and R-type sources show resolved double jet emission with a single visible central peak (ratio = 0; \citealt{yew22prep}).
The dataset is further expanded in \citet{gupta24b} with 2,100 compact/unresolved galaxies and 100 rare morphologies, including bent-tailed galaxies, cluster halo emissions, and Odd Radio Circles \citep[ORCs;][]{norris21b}. 
Gal-DINO refines bounding box and keypoint predictions for identifying radio sources and their infrared hosts. 
The performance evaluation yields an average precision with 50\% intersection over union (IoU), i.e., AP$_{50}$, of 73.2\% for bounding boxes and 71.7\% for keypoints, with 99\% of central bounding boxes achieving IoU > 0.5 and 98\% of keypoints located within $<3^{\prime \prime}$ of their true host positions \citep[see][]{gupta24b}.
We extend RG-CAT from EMU PS1 to the first-year EMU main survey tiles, generating $8^{\prime} \times 8^{\prime}$ cutouts for approximately 3 million \textit{Selavy}-based sources. 
Each cutout is analysed with Gal-DINO to extract bounding boxes, categories, and confidence scores, assembling a catalogue per tile. 
Compact sources are catalogued individually, while extended galaxies are grouped. 
A detailed catalogue of radio sources and host galaxies will be presented in \citet{gupta25bprep}, while this study focuses on extended radio sources including rare morphologies.

\section{Foundation Models and Fine-tuning}
\label{SEC:model}
Foundation models capture broad, transferable knowledge and can be fine-tuned to perform specific tasks in astronomy using relatively small amounts of labelled data. In this work, we fine-tune OpenCLIP, a multimodal foundation model, using radio source images and their corresponding textual descriptions. This enables the model to learn the unique visual and semantic features of radio sources. As a result, it can support downstream tasks such as retrieving similar images based on a query image or a text prompt. In this section, we discuss multimodal foundation models and provide details on fine-tuning OpenCLIP for the radio source dataset.
Figure~\ref{FIG:EMUSE} provides an overview of our framework.

\subsection{Multimodal Foundation Models}
\label{SEC:model1}
Foundation models have recently gained significant attention for their ability to integrate and process multiple modalities, such as images and text, within a unified framework. Multimodal image-text foundation models, in particular, have demonstrated remarkable capabilities in bridging the gap between vision and language, enabling applications like image captioning and visual question answering \citep[e.g.,][]{ramesh22,rombach22}. These models are typically pre-trained on large-scale datasets containing paired image-text data, such as captions or descriptions, using self-supervised learning techniques \citep[e.g.,][]{wang2021simvlm,cherti23}.
The self-supervised training paradigm leverages the inherent alignment between images and their corresponding textual descriptions to learn rich, joint representations without requiring explicit human annotations for every task. For instance, models like CLIP \citep[Contrastive Language–Image Pre-training;][]{radford21} and ALIGN \citep[][]{jia21} employ contrastive learning objectives, where the model learns to maximise the similarity between embeddings of matching image-text pairs while minimising it for non-matching pairs.

In contrast, GPT-based multimodal models extend the autoregressive language modelling paradigm of GPT to incorporate visual inputs \citep[e.g.,][]{alayrac2022flamingo}. These models are trained to predict the next token in a sequence, enabling them to generate coherent text conditioned on both textual and visual inputs. Unlike CLIP, which focuses on alignment, GPT-based models emphasise the generation of text based on multimodal inputs.
Gemini represents a unified architecture that aims to seamlessly integrate multiple modalities into a single cohesive model \citep[][]{gemini23}. Unlike CLIP, which separates vision and language encoders, and GPT-based models, which primarily extend language models to handle visual inputs, Gemini is designed to natively process multiple modalities (e.g., text, images, audio, video) within a single architecture.
Similarly, models like MultiMAE \citep[Multi-modal Multi-task Masked Autoencoders][]{bachmann2022multimae} use masked reconstruction tasks, where parts of the input (e.g., patches of an image or words in a sentence) are masked, and the model is trained to reconstruct them based on the remaining context.

\begin{figure}[htbp]
\centering
\includegraphics[width=8.2cm, scale=0.5]{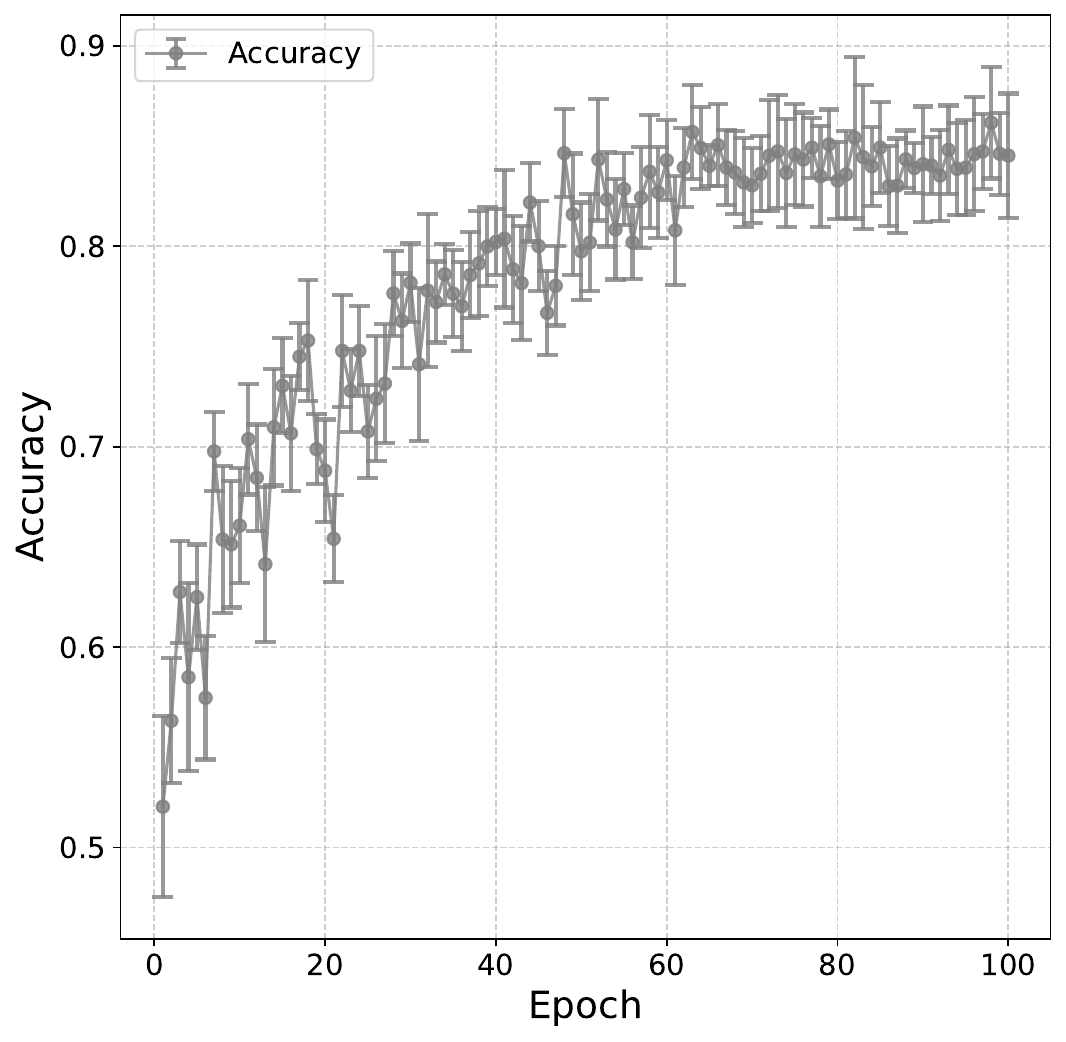}
\caption{Model accuracy evaluated on the test set after each epoch. Error bars represent the variance, calculated by fine-tuning and testing the model 10 times with randomly drawn training and test sets.} 
\label{FIG:finetune_valid}
\end{figure}

\subsection{Fine-tuning Foundation Model}
\label{SEC:model2}
The success of multimodal image-text foundation models lies in their ability to generalise across diverse tasks and domains by leveraging the complementary information in both modalities. 
By pre-training on vast amounts of image-text pairs, these models capture intricate cross-modal relationships, enabling them to excel in downstream tasks with minimal fine-tuning \citep[e.g.,][]{yu22,cherti23,touvron23}. 
Furthermore, the self-supervised nature of their training allows them to scale effectively with increasing data and computational resources, leading to emergent capabilities such as zero-shot or few-shot generalisation \citep[e.g.,][]{bommasani21,jia21,wang2021simvlm,alayrac2022flamingo}. 
Despite their successes, several challenges persist. These include the need for high-quality, diverse datasets for pre-training and the substantial computational resources required to train and deploy large-scale models.
The limited availability of open-source multimodal foundation models has also hindered their adoption in specialised fields like astronomy. 
However, recent collaborative efforts have led to the release of open-source multimodal pre-trained models, making them accessible to the broader research community. 

In this study, we use the OpenCLIP \citep[][]{cherti23}, an open-source multimodal foundation model, trained on 2.32 billion real-world image-text pairs sourced from the publicly accessible LAION dataset \citep[][]{schuhmann22laion}.
OpenCLIP is based on the CLIP architecture \citep[][]{radford21}.
OpenCLIP employs a Contrastive-Captioning \citep[CoCa;][]{yu22} framework that combines contrastive learning and generative captioning into a single unified model.
Contrastive learning aligns image and text embeddings in a shared latent space.
Generative captioning produces descriptive captions for images.
This dual-objective approach allows OpenCLIP to serve as a strong foundation model for both discriminative and generative multimodal tasks.
LAION is one of the largest open datasets for vision-language research, containing diverse and noisy web-scraped data that enable the model to learn robust cross-modal representations. 
By leveraging this vast amount of paired data, OpenCLIP achieves strong performance across a variety of tasks, including zero-shot image classification, cross-modal retrieval, and visual question answering. 
Fine-tuning OpenCLIP for specific downstream tasks is facilitated by its modular architecture and compatibility with widely used deep learning frameworks such as PyTorch.
Users can refine the model by updating all parameters or employing parameter-efficient approaches, such as linear probing or adapter-based fine-tuning.
In linear probing, only a task-specific classification head is trained while keeping the pre-trained weights fixed. 
This makes it a computationally efficient strategy, particularly for applications with limited labelled data.
For more complex tasks, full fine-tuning enables the model to adapt its learned representations to the specific characteristics of the target domain.
Furthermore, OpenCLIP allows for customisation through modifications to its training pipeline, providing flexibility to explore alternative objectives, optimisers, and data augmentation techniques.

We use the RadioGalaxyNET dataset \citep[][]{gupta24a} to fine-tune the pre-trained OpenCLIP model.
The dataset includes 2,800 FR-I, FR-II, FR-x, and R-type radio galaxies, along with their corresponding infrared hosts.
Following \citep[][]{gupta24b}, we incorporate an additional category containing 100 peculiar sources and other rare morphologies.
For each of these radio sources, we generate $4^{\prime} \times 4^{\prime}$ image cutouts from the EMU-PS1 survey and corresponding cutouts from the AllWISE survey. 
The host galaxy position is used as the cutout centre, ensuring that the full extent of the radio emission is captured.
These cutouts are saved as PNG (Portable Network Graphics) images, with the first two channels containing radio cutouts. Data clipping is applied between the 50th percentile level and the maximum values of the 99th and 99.9th percentiles for the first and second channels, respectively. 
The third channel contains the AllWISE W1 band image.
We expand the labels for these radio galaxies by incorporating morphological descriptions and textual variations (see examples in Table~\ref{TAB:expanded_text}), and by adding additional information based on their subcategories \citep{yew22prep}.
For instance, an FR-II radio galaxy that exhibits a bent-tailed structure is labelled as:
``An image of an FR-II or Fanaroff-Riley type II radio galaxy with edge-brightened lobes bent at an angle."
Similarly, an ORC, an extragalactic, edge-brightened ring-like radio structure surrounding a distant host galaxy, typically lacks detectable emission at other wavelengths beyond its host but can exhibit diffuse radio emission within the bright ring structure \citep[][]{norris24}, and is labelled as:
``An image of a peculiar radio galaxy classified as an Odd Radio Circle."
Additional sub-categories include HyMORS (hybrid morphology radio sources), which exhibit an FR-I appearance on one side of the core and an FR-II appearance on the other; DDRGs (double-double radio galaxies), often interpreted as "restarted" radio galaxies; resolved star-forming radio galaxies; as well as core-dominated radio galaxies where the radio emission associated with the host galaxy is significantly brighter than the lobes.

\begin{figure}[!ht]
\centering
\includegraphics[trim=0 0 1.4cm 0, width=9cm, scale=0.5]{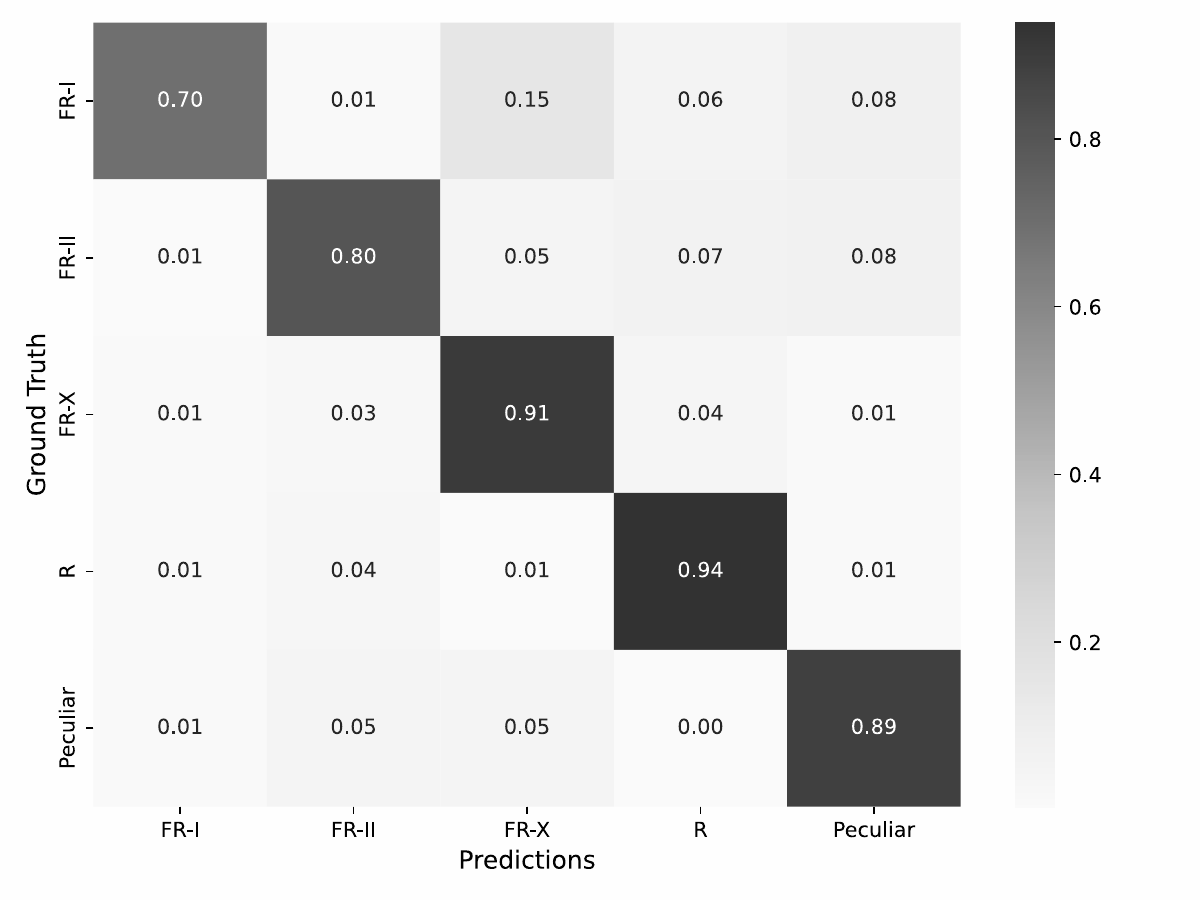}
\includegraphics[trim=0 0 0 0, width=8.5cm, scale=0.5]{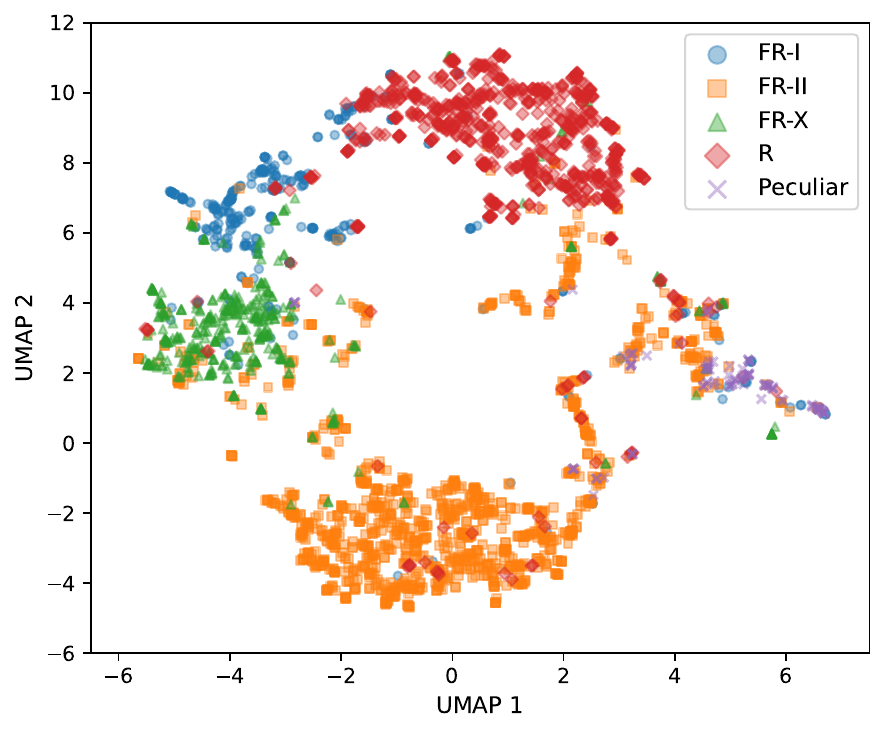}
\caption{The top panel shows the confusion matrix comparing ground truth labels to predicted labels for each main category. The displayed values are averaged over 10 training iterations. The bottom panel shows the UMAP projection generated from the image embeddings produced by the model's image encoder, illustrating that different ground truth categories cluster in distinct regions. The plotted points include test sets from all 10 training iterations.} 
\label{FIG:confusion_mat}
\end{figure}

Using the radio and infrared image cutouts of sources along with the expanded text descriptions, we fine-tune the pre-trained OpenCLIP model on a single NVIDIA H100 GPU for 100 epochs, which takes approximately 1.5 hours.
We employ adapter-based fine-tuning, which allows the model to adapt its learned representations to the characteristics of radio sources.
Given that OpenCLIP combines both the contrastive and generative sides into a single unified architecture, we focus solely on the contrastive side during fine-tuning. 
This approach encourages embeddings of matching image-text pairs to be close together while pushing non-matching pairs apart, thereby enabling zero-shot retrieval tasks for EMU data.
To evaluate the model's performance, we split the radio source dataset into an 80:20 ratio for training and testing. The training and testing data are randomly sampled from the full set 10 times, and the OpenCLIP model is trained separately on each iteration of the randomly selected training data. The trained models are then tested on independently selected test data, also drawn randomly 10 times.
Figure~\ref{FIG:finetune_valid} presents the accuracy over 100 training epochs. The error bars reflect the variance in test results across the 10 training iterations. The figure indicates that accuracy exceeds 50\% after a single epoch and gradually increases to $84\pm3$\% after 100 epochs.
Notably, while the model is trained on images paired with their expanded text descriptions, we assess its accuracy using only the main categories -- FR-I, FR-II, FR-x, R, and Peculiar -- during testing. 
Top panel of Figure~\ref{FIG:confusion_mat} shows the confusion matrix for these main categories. The values shown are averaged across 10 training iterations. The results demonstrate that the fine-tuned model predicts these categories with high accuracy overall, although there is greater confusion between FR-I and FR-x sources. This is expected, as the primary distinction between these two categories lies in the peak-to-extent ratio (as described in Section~\ref{SEC:dataset3}). In contrast, confusion is much lower for the Peculiar category, despite it having the smallest training sample size.
Bottom panel of Figure~\ref{FIG:confusion_mat} displays the Uniform Manifold Approximation and Projection \citep[UMAP,][]{mcinnes18} projection of image embeddings from the model, with points representing sources in test sets across all 10 training runs. This highlights how different ground truth categories form distinct clusters, while also revealing overlaps that align with the patterns seen in the confusion matrix.
Additionally, although the accuracy and confusion matrix evaluations are based on training with 80\% of the data, we fine-tune the final model using 100\% of the radio source dataset. This ensures that all available image-text pairs are utilised to train the final model used for the EMU search engine.

\section{EMUSE Application}
\label{SEC:emuse}
We develop EMUSE (Evolutionary Map of the Universe Search Engine), a tool that employs similarity search using the fine-tuned model described in the previous section.
We use catalogues generated by the RG-CAT pipeline (see Section~\ref{SEC:dataset3}), which employs the Gal-DINO object detection model to process each EMU tile.
We filter extended radio sources classified as FR-I, FR-II, FR-x, R, and Peculiar from the catalogues.
From the 160 tiles observed during the first year of the EMU survey, we identify approximately 170,000 such extended radio sources where the prediction confidence score exceeds the minimum estimated threshold of the Gal-DINO model.
Using the sky positions from the catalogues, we generate cutouts from the EMU and AllWISE surveys, which are saved as radio-radio-infrared channel PNG images.
The fine-tuned model is then used to generate image embeddings for each PNG.
Additionally, we store the corresponding catalogue metadata for each image embedding, including source positions, integrated radio flux, and the potential host name from the CatWISE catalogue \citep[][]{marocco21}, as provided by the RG-CAT pipeline.
Note that the potential host details provided here are based on estimates from the Gal-DINO model within the RG-CAT pipeline and have not been verified through visual inspection.

EMUSE implements a zero-shot retrieval framework, enabling the model to generalise its knowledge to unseen classes or tasks without explicit training on those specific classes.
In this work, we use the fine-tuned OpenCLIP multimodal model, which has been trained to produce aligned embeddings for images and text.
Specifically, we generate embeddings for approximately 170,000 EMU survey radio sources from PNGs with radio and infrared channels, using the fine-tuned model.
These embeddings replace the original images, which require over 150 GB of storage and are difficult to search efficiently for multiple queries. 
In contrast, the embeddings occupy only a few hundred megabytes, making the search engine viable.
These embeddings are stored in a database and can be queried using either text queries (e.g., ``radio galaxy with jets'') or image queries (e.g., a sample image of a radio source).
The zero-shot capability arises from the model's ability to retrieve similar sources based on the semantic alignment of embeddings in the shared latent space, without requiring additional training on specific classes or queries.

For a given text query, the input is first tokenised using the OpenCLIP tokeniser, and its embedding is obtained through the fine-tuned model's text encoder. For an image query, the input image undergoes preprocessing using OpenCLIP's standard pipeline, which includes resizing to $224\times224$ pixels, conversion to RGB and then to Pytorch tensor, and normalisation with the model's predefined mean and standard deviation values. The resulting image is then passed through the fine-tuned model's image encoder to generate its embedding.
To search for similar sources, we compute the similarity between the query embedding (either derived from a text or an image query) and the precomputed embeddings of the EMU survey source images as

\begin{equation} 
S(\mathbf{q}, \mathbf{e}_i) = \frac{\mathbf{q} \cdot \mathbf{e}_i}{||\mathbf{q}||~||\mathbf{e}_i||}, 
\end{equation}

where:

\begin{itemize} 
\item $ \mathbf{q} \in \mathbb{R}^d $: The embedding of the query (text or image) in the shared latent space.
\item $ \mathbf{e}_i \in \mathbb{R}^d $: The embedding of the $ i $-th image in the database ($ i = 1, 2, \dots, N $).
\item $ S(\mathbf{q}, \mathbf{e}_i) $: The cosine similarity function measures the alignment between the query and image embeddings, normalised between 0 and 1.
\end{itemize}

The top-$k$ most similar image embeddings are retrieved as

\begin{equation}
\text{top-}k = \argmax_{i \in \{1, 2, \dots, N\}} S(\mathbf{q}, \mathbf{e}_i).
\end{equation}

The information corresponding to these top-$k$ embeddings is then fetched from the RG-CAT catalogue metadata. 
This includes the EMU tile SBID where the source is located, its RA ($\deg$), Dec ($\deg$), integrated flux density (mJy), and potential host galaxy names from the CatWISE catalogue, along with the probability describing the estimated similarity between the query embedding $\mathbf{q}$ and the image embedding $\mathbf{e}_i$. 
The following sections discuss examples of text and image queries.

\subsection{Text Queries}
We evaluate the zero-shot retrieval capability of the fine-tuned OpenCLIP model using various queries, presenting two examples for brevity. 
The application is publicly available, allowing readers to submit their queries.
For instance, we search for ``A bent-tailed radio galaxy". Table~\ref{TAB:bent_tailed} displays the EMUSE output, listing the top 50 most similar radio sources along with their potential host galaxies from RG-CAT. 
The number of displayed sources can be adjusted by modifying the minimum probability threshold and the desired number of results in the interface. 
Using the positions in Table~\ref{TAB:bent_tailed}, we present all 50 corresponding images in Figure~\ref{FIG:bent_tailed}, demonstrating that the fine-tuned model can efficiently retrieve bent-tailed radio sources across the EMU survey.
For the second query, ``Resolved star-forming radio galaxies", the EMUSE results are shown in Table~\ref{TAB:starforming} and Figure~\ref{FIG:starforming}, further highlighting the model's ability to identify and classify such morphologies.
While these examples showcase the model’s capability to interpret text queries and retrieve relevant image data, this performance is directly attributed to the fine-tuning applied in this work. 
Sources absent from the fine-tuning dataset -- such as cluster relics and supernova remnants -- may not be retrieved effectively.

\begin{figure}[htbp]
\centering
\includegraphics[width=4cm, height=3.59cm]{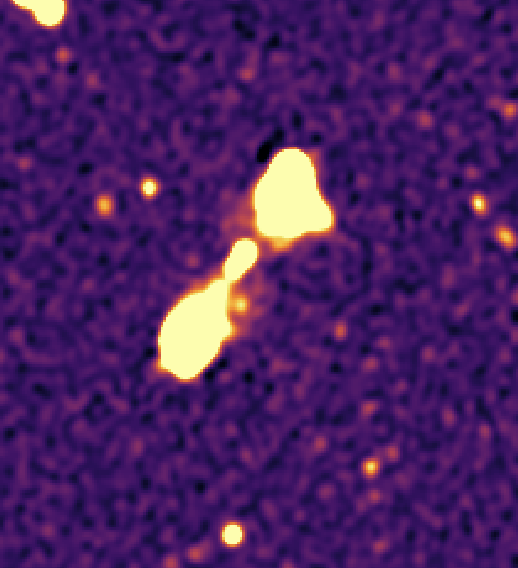}
\includegraphics[width=4cm, scale=0.5]{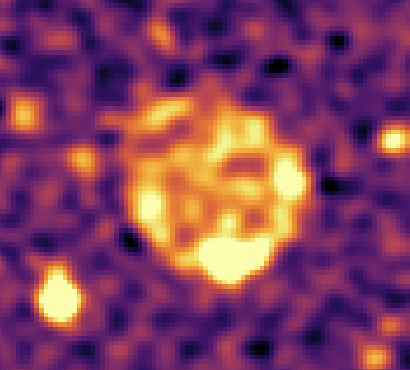}
\caption{Example image queries for EMUSE. 
These figures are screenshots from the EMU-PS1 image, taken while being viewed in CARTA. 
The left panel shows an FR-II radio galaxy, while the right panel displays ORC J2103-6200 \citep{norris21b}.
}
\label{FIG:imagequeries}
\end{figure}

Additionally, text-based queries in EMUSE currently underperform compared to image-based queries. For example, a simple search for ``odd radio circle'' returns no results above a probability threshold of 0.9, while a more descriptive prompt, such as ``An image of a peculiar radio galaxy classified as an Odd Radio Circle'', successfully retrieves relevant sources. Conversely, concise text like ``FR-II'' yields meaningful matches, whereas longer, more complex phrases, such as ``An image of an FR-II or Fanaroff-Riley type II radio galaxy with edge-brightened lobes bent at an angle'', often result in inconsistent or unrelated outputs. This inconsistency stems from the sensitivity of the model to phrasing and its reliance on the limited and sparse textual descriptions used during fine-tuning. Since the alignment between text and image embeddings depends heavily on how descriptions are written, the model struggles to interpret astronomy-specific language without sufficient contextual variety. While adding a broader range of textual descriptions could help, this approach is constrained by variability in human annotation styles. A more scalable and effective solution may involve augmenting the training data with language rewrites \citep{fan23} and paraphrasing techniques \citep{hyunjae24} or by leveraging large language models to generate richer and more diverse textual descriptions \citep[e.g.,][]{nguyen23,yu24,chen24}. These strategies could enhance the model’s ability to interpret different forms of scientific language and better align them with corresponding visual features, and should be explored in future work.

\subsection{Image Queries}
For image-based queries, we demonstrate two examples: an FR-II radio galaxy and ORC J2103-6200 \citep[][]{norris21b}. 
We use EMU-PS1 images, open them in CARTA\footnote{\url{https://cartavis.org/}}, and capture screenshots of these sources (see Figure~\ref{FIG:imagequeries}). 
These screenshots are then used as query inputs to search the EMU survey.
For the FR-II source shown in the left panel of Figure~\ref{FIG:imagequeries}, the corresponding EMUSE results are presented in Table~\ref{TAB:fr2} and Figure~\ref{FIG:fr2}. 
Notably, most of the retrieved sources exhibit emission from the core, which is consistent with the query image. 
Additionally, their sky orientation closely matches that of the input query, further demonstrating the model's effectiveness in retrieving morphologically similar sources.

The EMUSE results for the ORC J2103-6200 image query are shown in Table~\ref{TAB:orc1}, and in Figure~\ref{FIG:orc1}. 
The first four sources include a starburst radio ring galaxy (SRRG), an ORC candidate, another SRRG, and a radio source without a plausible host galaxy, as also identified in \citet{gupta25a}.
Although the training set for fine-tuning included only two ORCs, the model successfully retrieves a known ORC candidate, several half-ring-like structures, and potential GLAREs (Galaxies with Large-scale Ambient Radio Emission; \citealt{gupta25a}), which may represent an evolutionary stage of ORCs.
This demonstrates the potential of EMUSE for discovering such rare radio sources, which will be enhanced by incorporating a larger training sample of these sources in future updates to the model.
Further multi-wavelength visual inspections are needed to categorise the remaining sources in the figure. 
Due to the limited training data for ORCs, the model also retrieves resolved star-forming radio galaxies and other radio sources occupying similar embedding spaces to the image query. 
However, it also identifies Wide Angle Tailed (WAT) sources and other diffuse emissions, highlighting the need for more ORC examples in the training data.

Note that when a screenshot is used as a query input to a model trained on 3-channel images, the information in the image is typically replicated across all three channels to match the expected input format. Although the screenshot may lack the multi-channel radio and infrared details present in the training data, the model often still performs reasonably well. This is likely because high-level structural features, such as morphology and spatial patterns, are still available. While the resulting embeddings may not capture the full richness of the original data, such as distinguishing between resolved spirals and ORCs, they can still yield meaningful similarity results.
Additionally, we find that different image queries -- such as screenshots of this ORC taken from various sources (e.g., academic papers) or images of other previously identified ORCs and ORC candidates -- yield different sets of sources in the similarity space. 
A comprehensive future study of similar sources obtained from various queries will help expand the catalogue of such rare systems.

\section{Conclusions}
\label{SEC:conclusions}
We explore the application of multimodal foundation models in the field of radio astronomy, specifically leveraging the power of OpenCLIP, an open-source pre-trained multimodal model, to classify and retrieve radio sources from the EMU survey. 
Radio astronomy, with its vast and complex datasets, benefits from advanced machine learning techniques that can efficiently process large amounts of data and provide insights into the nature of celestial objects. 
This paper aims to enhance the identification and retrieval of different types of radio galaxies by using the OpenCLIP model, which integrates both visual and textual information in a shared embedding space. 
The motivation behind this study is to bridge the gap between machine learning and astronomy, allowing for more accurate and efficient searches within large radio source databases.

In this work, we fine-tune the OpenCLIP model on a dataset of 2,900 radio galaxies from the RadioGalaxyNET dataset, which includes various morphological classes, such as FR-I, FR-II, FR-x, R-type, and peculiar radio sources. 
The fine-tuning is performed using adapter-based methods, ensuring that the model adapts effectively to the specific characteristics of radio sources while maintaining computational efficiency. 
The model is trained to map radio and infrared images to a shared latent space alongside their associated textual descriptions. 
Through this process, the model learns the complex relationships between image features and text, making it capable of performing zero-shot retrieval tasks without the need for additional task-specific training.

The fine-tuned OpenCLIP model is then integrated into the EMUSE (Evolutionary Map of the Universe Search Engine) application, enabling the efficient search and retrieval of radio sources from the EMU survey. 
By converting the images of radio sources into compact embeddings, the model reduces the data storage requirements and makes searching across large datasets feasible. 
The application allows users to query the database using both text and image-based inputs, providing a flexible and powerful tool for identifying and classifying radio galaxies. 
Notably, the zero-shot retrieval capabilities of the model allow it to generalise to new types of radio sources, making it adaptable to future discoveries without the need for retraining.

The results from the evaluation of the model demonstrate its effectiveness in retrieving radio sources based on both text and image queries. In particular, the model performs well in retrieving sources with specific morphological features. 
Additionally, the image query functionality highlights the model’s ability to recognise and retrieve similar sources with matching morphological features, even for complex objects like Odd Radio Circles. 
However, certain categories of radio sources--such as supernova remnants, planetary nebulae, cluster relics, etc.--which were absent from the fine-tuning dataset may not be retrieved as accurately.
This limitation highlights the importance of continuously expanding the training data to include a wider range of radio source types.

Future work should focus on extending the model to accommodate more complex datasets, enhancing its performance on rare or previously unseen radio sources, and integrating it with other astronomical databases to further expand its capabilities. 
Future work should also focus on improving the accessibility of the EMUSE application by displaying the source images from the catalogue generated through image and text queries. This functionality can be implemented by retrieving images via the cutout service, which is currently being integrated into the CASDA server.
While this study demonstrates the model’s application using the first-year data from the EMU survey, future efforts should incorporate observations from the ongoing survey in the coming years.
In addition, incorporating more multiwavelength datasets will help refine the classification of rare radio sources, improving the model’s accuracy and applicability. 
The current approach relies on RG-CAT catalogues, which in turn are derived from {\it Selavy}-based catalogues. 
Consequently, sources missed by {\it Selavy}--such as very faint objects--are also absent from our results. 
Future research should explore catalogue-agnostic approaches to mitigate this limitation.
Furthermore, with the increasing availability of open-source pre-trained models, whether trained on astronomical or real-world data, future studies should investigate the adoption of newer architectures that may enhance fine-tuning beyond OpenCLIP.
By providing an efficient and scalable solution for radio astronomy, this approach paves the way for researchers to explore and classify the ever-growing volume of radio data more effectively, ultimately advancing our understanding of complex radio sources.

\section*{Data Availability}
The OpenCLIP model with fine-tuning settings is available at \url{https://github.com/Nikhel1/Finetune_OpenCLIP}. 
The radio source images and labels used for fine-tuning are available at \url{https://doi.org/10.25919/btk3-vx79}, while the exact images and expanded text descriptions are available upon request.
The search engine is accessible at \url{https://askap-emuse.streamlit.app/} and can also be used locally by cloning the repository and following the steps provided at \url{https://github.com/Nikhel1/EMUSE}, i.e., by running the command ``streamlit run main.py".
The fine-tuned models, EMU survey radio source embeddings, and catalogue metadata are accessible within ``main.py".

\section*{Acknowledgements}
NG acknowledges support from CSIRO’s Machine Learning and Artificial Intelligence Future Science Impossible Without You (MLAI FSP IWY) Platform.
This scientific work uses data obtained from Inyarrimanha Ilgari Bundara / the Murchison Radio-astronomy Observatory. We acknowledge the Wajarri Yamaji People as the Traditional Owners and native title holders of the Observatory site. The Australian SKA Pathfinder is part of the Australia Telescope National Facility (https://ror.org/05qajvd42) which is managed by CSIRO. Operation of ASKAP is funded by the Australian Government with support from the National Collaborative Research Infrastructure Strategy. ASKAP uses the resources of the Pawsey Supercomputing Centre. The establishment of ASKAP, the Murchison Radio-astronomy Observatory and the Pawsey Supercomputing Centre are initiatives of the Australian Government, with support from the Government of Western Australia and the Science and Industry Endowment Fund.
This paper includes archived data obtained through the CSIRO ASKAP Science Data Archive, CASDA (\url{http://data.csiro.au}).

This publication makes use of data products from the Wide-field Infrared Survey Explorer, which is a joint project of the University of California, Los Angeles, and the Jet Propulsion Laboratory/California Institute of Technology, and NEOWISE, which is a project of the Jet Propulsion Laboratory/California Institute of Technology. WISE and NEOWISE are funded by the National Aeronautics and Space Administration.

We acknowledge the use of several open-source Python packages that facilitated this research, including (but not limited to) PyTorch \citep{pytorch}, scikit-learn \citep{scikit-learn}, pandas \citep{pandas}, and Astropy \citep{astropy:2013, astropy:2018, astropy:2022}.

\bibliography{ASKAP_PASA}

\appendix
\renewcommand{\thetable}{A\arabic{table}} 
\renewcommand{\thefigure}{A\arabic{figure}} 
\setcounter{figure}{0} 


\begin{table*}[!htbp]
    \centering
    \begin{tabular}{c c c c c c}
    \hline
    \textbf{SBID} & \textbf{RA (Degrees)} & \textbf{Dec (Degrees)} & \textbf{Integrated Flux (mJy)} & \textbf{CatWISE Potential Host} & \textbf{Probability} \\
    \hline
    47034 & 37.63730 & -49.23537 & 8.27 & J023032.95-491407.3 & 0.99 \\
    53218 & 251.15865 & -61.97237 & 15.71 & J164438.07-615820.5 & 0.99 \\
    51432 & 151.54313 & -10.53717 & 23.71 & J100610.35-103213.7 & 0.99 \\
    50419 & 137.03420 & -5.72172 & 31.96 & J090808.20-054318.1 & 0.99 \\
    51964 & 335.60979 & -4.71453 & 51.79 & J222226.34-044252.3 & 0.99 \\
    59804 & 138.76081 & -16.81240 & 9.01 & J091502.59-164844.6 & 0.99 \\
    46982 & 47.54216 & -69.89382 & 14.76 & J031010.11-695337.7 & 0.99 \\
    51962 & 209.24932 & -8.51805 & 7.80 & J135659.83-083104.9 & 0.99 \\
    59095 & 151.54427 & -10.53891 & 25.03 & J100610.62-103220.0 & 0.99 \\
    51852 & 329.42790 & -4.31361 & 6.29 & J215742.69-041848.9 & 0.99 \\
    54923 & 330.48922 & -62.24350 & 5.05 & J220157.41-621436.6 & 0.99 \\
    52145 & 115.69195 & -56.67697 & 261.32 & J074246.06-564037.0 & 0.99 \\
    59607 & 56.37160 & -72.79562 & 6.41 & nan & 0.99 \\
    53557 & 355.02405 & -69.47579 & 60.33 & nan & 0.99 \\
    54098 & 118.43152 & -48.97921 & 8.79 & J075343.56-485845.1 & 0.99 \\
    50417 & 249.27072 & -73.72866 & 5.27 & J163704.97-734343.1 & 0.98 \\
    54802 & 293.68269 & -68.72743 & 12.80 & J193443.84-684338.7 & 0.98 \\
    54770 & 77.56348 & -9.15113 & 31.80 & J051015.23-090904.0 & 0.98 \\
    59609 & 181.47904 & 0.76041 & 0.93 & J120554.96+004537.4 & 0.98 \\
    51559 & 197.17494 & -4.94096 & 87.75 & J130841.98-045627.4 & 0.98 \\
    47034 & 41.25039 & -50.51419 & 40.77 & J024500.09-503051.0 & 0.98 \\
    50534 & 171.09989 & -0.74536 & 1.49 & J112423.97-004443.2 & 0.98 \\
    52161 & 332.68165 & -10.94630 & 0.63 & J221043.59-105646.6 & 0.98 \\
    53566 & 334.34896 & -71.74171 & 2.21 & nan & 0.98 \\
    53218 & 251.11829 & -61.97387 & 1.25 & nan & 0.98 \\
    46971 & 13.11986 & -37.73006 & 47.53 & J005228.76-374348.2 & 0.98 \\
    50181 & 133.86841 & 1.80529 & 2.29 & J085528.41+014819.0 & 0.98 \\
    55325 & 333.19754 & -8.31449 & 2.45 & J221247.41-081852.1 & 0.98 \\
    46971 & 10.68130 & -38.09499 & 36.30 & J004243.51-380541.9 & 0.98 \\
    45781 & 334.95910 & -57.38690 & 5.19 & J221950.18-572312.8 & 0.98 \\
    54105 & 266.01302 & -53.35766 & 5.58 & J174403.12-532127.5 & 0.98 \\
    46946 & 32.10990 & -53.65888 & 18.66 & nan & 0.98 \\
    46971 & 14.40059 & -34.83221 & 5.28 & J005736.14-344955.9 & 0.98 \\
    59609 & 181.08124 & -0.31712 & 6.30 & J120419.49-001901.6 & 0.98 \\
    54807 & 29.34617 & -9.45147 & 279.85 & J015723.08-092705.2 & 0.98 \\
    55325 & 335.34370 & -11.61309 & 3.50 & J222122.48-113647.1 & 0.98 \\
    52145 & 116.96675 & -57.52042 & 10.44 & J074752.02-573113.5 & 0.98 \\
    51962 & 208.93476 & -7.27899 & 4.04 & J135544.34-071644.3 & 0.98 \\
    46955 & 264.45051 & -71.44783 & 11.33 & nan & 0.98 \\
    55325 & 332.58552 & -10.21272 & 11.37 & J221020.52-101245.7 & 0.98 \\
    54926 & 265.19108 & -62.84665 & 1.35 & J174045.85-625047.9 & 0.97 \\
    50423 & 30.06071 & -28.35766 & 1.54 & J020014.56-282127.5 & 0.97 \\
    50787 & 52.00056 & -52.70249 & 0.37 & J032800.13-524208.9 & 0.97 \\
    51818 & 304.85137 & -56.85586 & 64.85 & J201924.32-565121.0 & 0.97 \\
    47034 & 34.02686 & -52.14808 & 1.38 & J021606.44-520853.0 & 0.97 \\
    52125 & 299.83536 & -59.08474 & 10.32 & J195920.48-590505.0 & 0.97 \\
    51930 & 314.36795 & 3.41040 & 7.30 & J205728.30+032437.4 & 0.97 \\
    46971 & 14.10576 & -39.69648 & 106.41 & nan & 0.97 \\
    51931 & 83.41189 & -57.68088 & 8.26 & J053338.85-574051.1 & 0.97 \\
    53210 & 102.14236 & -45.50372 & 8.87 & nan & 0.97 \\
    50049 & 130.24632 & 1.60090 & 29.77 & J084059.11+013603.2 & 0.97 \\
        \hline
    \end{tabular}
    \caption{Top-50 EMUSE output for text query, ``A bent-tailed radio galaxy".}
    \label{TAB:bent_tailed}
\end{table*}

\begin{figure*}[!htbp]
\centering
\includegraphics[width=0.98\textwidth]{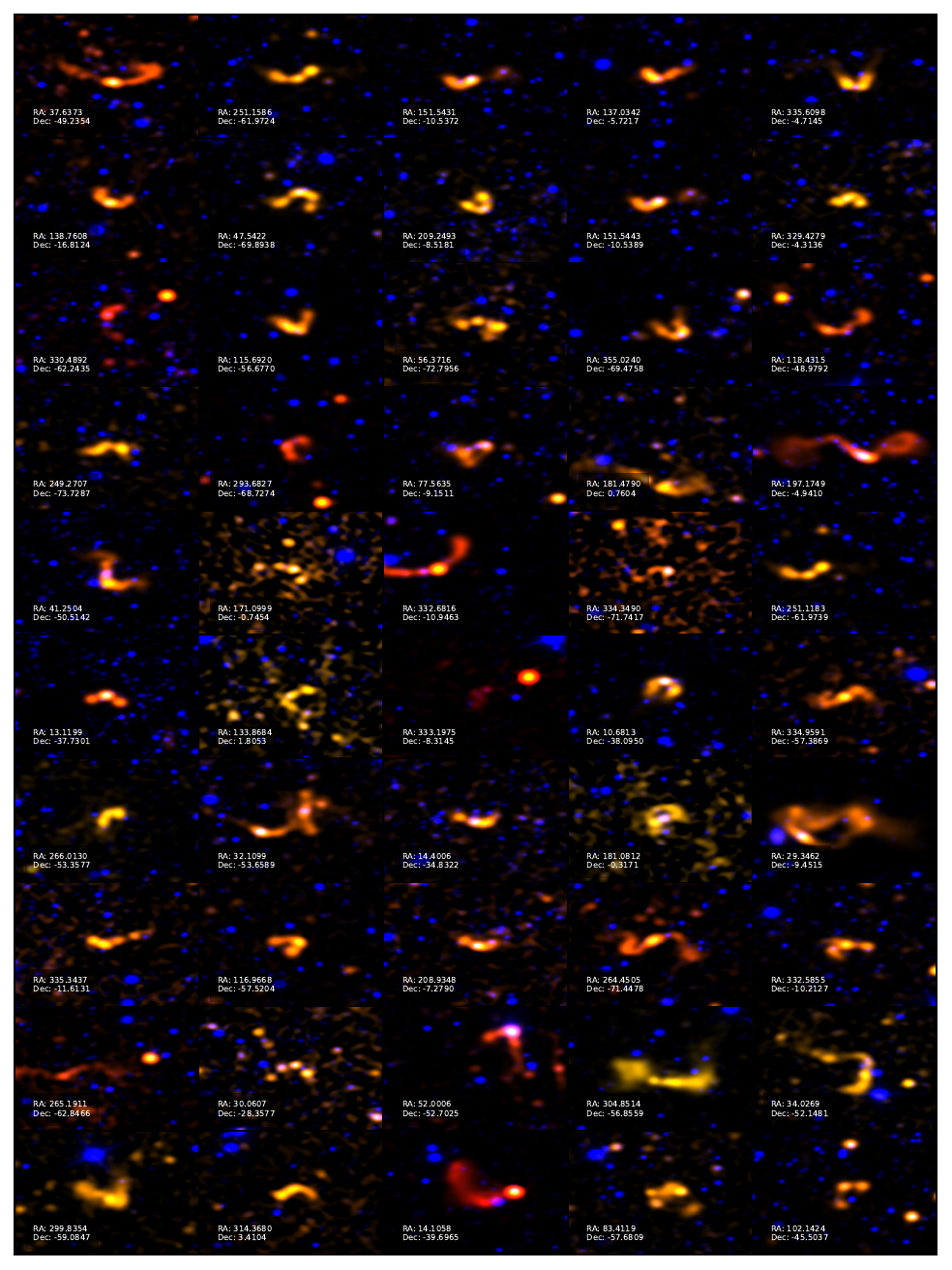}
\caption{Top-50 EMUSE output for the text query, ``A bent-tailed radio galaxy". Positions in Table~\ref{TAB:bent_tailed} are used here for $5^{\prime}\times5^{\prime}$ cutout images with radio-radio-infrared (RGB) channels.} 
\label{FIG:bent_tailed}
\end{figure*}

\begin{table*}[!htbp]
    \centering
    \begin{tabular}{c c c c c c}
    \hline
    \textbf{SBID} & \textbf{RA (Degrees)} & \textbf{Dec (Degrees)} & \textbf{Integrated Flux (mJy)} & \textbf{CatWISE Potential Host} & \textbf{Probability} \\
    \hline
    51958 & 340.06634 & -2.42481 & 6.32 & nan & 0.97 \\
    51964 & 340.06702 & -2.42499 & 4.63 & nan & 0.96 \\
    59560 & 180.09821 & -1.10006 & 212.74 & J120023.57-010600.2 & 0.96 \\
    54773 & 11.94711 & -11.46856 & 20.76 & J004747.30-112806.8 & 0.96 \\
    59607 & 68.25520 & -73.23706 & 54.31 & J043301.24-731413.4 & 0.96 \\
    52125 & 298.60649 & -58.71635 & 43.65 & nan & 0.96 \\
    59609 & 180.09846 & -1.09651 & 2.03 & nan & 0.95 \\
    51959 & 100.18053 & -58.52443 & 61.85 & J064043.32-583127.9 & 0.95 \\
    53513 & 340.10107 & -66.47889 & 1.77 & J224024.25-662844.0 & 0.95 \\
    51574 & 283.00855 & -57.32063 & 33.93 & J185202.05-571914.2 & 0.95 \\
    51932 & 197.20163 & -6.77416 & 34.15 & nan & 0.95 \\
    53313 & 3.62772 & -7.16736 & 23.08 & J001430.65-071002.4 & 0.94 \\
    46976 & 351.34005 & -57.79137 & 23.85 & J232521.61-574728.9 & 0.94 \\
    46925 & 35.01385 & -64.60222 & 49.33 & J022003.32-643607.9 & 0.94 \\
    59835 & 218.46783 & 5.45840 & 50.65 & J143352.27+052730.2 & 0.94 \\
    51574 & 278.63108 & -57.79277 & 39.97 & J183431.45-574733.9 & 0.94 \\
    51948 & 197.51854 & -46.43741 & 10.42 & J131004.45-462614.6 & 0.94 \\
    51930 & 319.81097 & 6.02001 & 4.54 & J211914.63+060112.0 & 0.93 \\
    52096 & 315.71100 & 3.94851 & 12.02 & nan & 0.93 \\
    54769 & 279.98512 & -67.42562 & 14.59 & J183956.42-672532.2 & 0.93 \\
    54770 & 74.19130 & -10.59294 & 8.26 & J045645.91-103534.5 & 0.93 \\
    50230 & 72.09941 & -59.80035 & 54.35 & J044823.85-594801.2 & 0.93 \\
    50182 & 210.52001 & -1.35790 & 6.31 & J140204.80-012128.4 & 0.93 \\
    50787 & 57.32923 & -51.81883 & 9.22 & J034919.01-514907.7 & 0.93 \\
    53314 & 113.84786 & -66.35405 & 4.08 & J073523.48-662114.5 & 0.93 \\
    46978 & 74.65443 & -75.07876 & 3.33 & nan & 0.93 \\
    52121 & 311.94228 & -65.08420 & 7.43 & J204746.14-650503.1 & 0.92 \\
    51948 & 197.50475 & -46.44542 & 0.75 & nan & 0.92 \\
    50230 & 73.21743 & -59.74236 & 40.85 & J045252.18-594432.5 & 0.92 \\
    54926 & 271.28683 & -64.19840 & 0.78 & J180508.83-641154.2 & 0.92 \\
    54802 & 295.97530 & -70.63307 & 126.75 & J194354.07-703759.0 & 0.92 \\
    52125 & 293.64487 & -61.14600 & 4.34 & J193434.77-610845.5 & 0.92 \\
    45761 & 328.32113 & -59.49363 & 26.04 & J215317.07-592937.0 & 0.92 \\
    46946 & 29.41844 & -57.79017 & 48.72 & J015740.42-574724.6 & 0.92 \\
    53566 & 330.52304 & -71.08281 & 2.61 & nan & 0.92 \\
    46951 & 16.25826 & -49.41661 & 165.21 & J010501.98-492459.7 & 0.92 \\
    53211 & 219.57554 & 3.41044 & 13.33 & J143818.13+032437.5 & 0.91 \\
    53566 & 346.08974 & -71.48710 & 18.93 & nan & 0.91 \\
    51931 & 83.01161 & -56.35367 & 45.12 & J053202.78-562113.2 & 0.91 \\
    54098 & 114.32972 & -52.74191 & 3.86 & J073719.13-524430.8 & 0.91 \\
    46955 & 274.10770 & -71.58137 & 19.48 & J181625.84-713452.9 & 0.91 \\
    59246 & 185.17599 & -0.86446 & 57.97 & J122042.23-005152.0 & 0.91 \\
    46980 & 259.24811 & -62.82057 & 29.04 & J171659.54-624914.0 & 0.91 \\
    47130 & 36.12181 & -44.60664 & 8.49 & J022429.23-443623.9 & 0.91 \\
    51928 & 202.98277 & -6.64210 & 32.91 & J133155.86-063831.5 & 0.90 \\
    59612 & 327.96281 & -69.08918 & 2.55 & J215151.07-690521.0 & 0.90 \\
    51930 & 315.71009 & 3.94843 & 12.21 & nan & 0.90 \\
    52125 & 295.89349 & -58.65578 & 181.12 & J194334.43-583920.8 & 0.90 \\
    53313 & 7.80539 & -10.48083 & 22.45 & J003113.29-102850.9 & 0.90 \\
    54923 & 319.06100 & -64.46230 & 4.87 & nan & 0.90 \\
    51948 & 199.74997 & -47.90849 & 15.66 & J131859.99-475430.5 & 0.90 \\
    \hline
    \end{tabular}
    \caption{Top-50 EMUSE output for text query, ``Resolved star forming radio galaxy".}
    \label{TAB:starforming}
\end{table*}

\begin{figure*}[!htbp]
\centering
\includegraphics[width=0.98\textwidth]{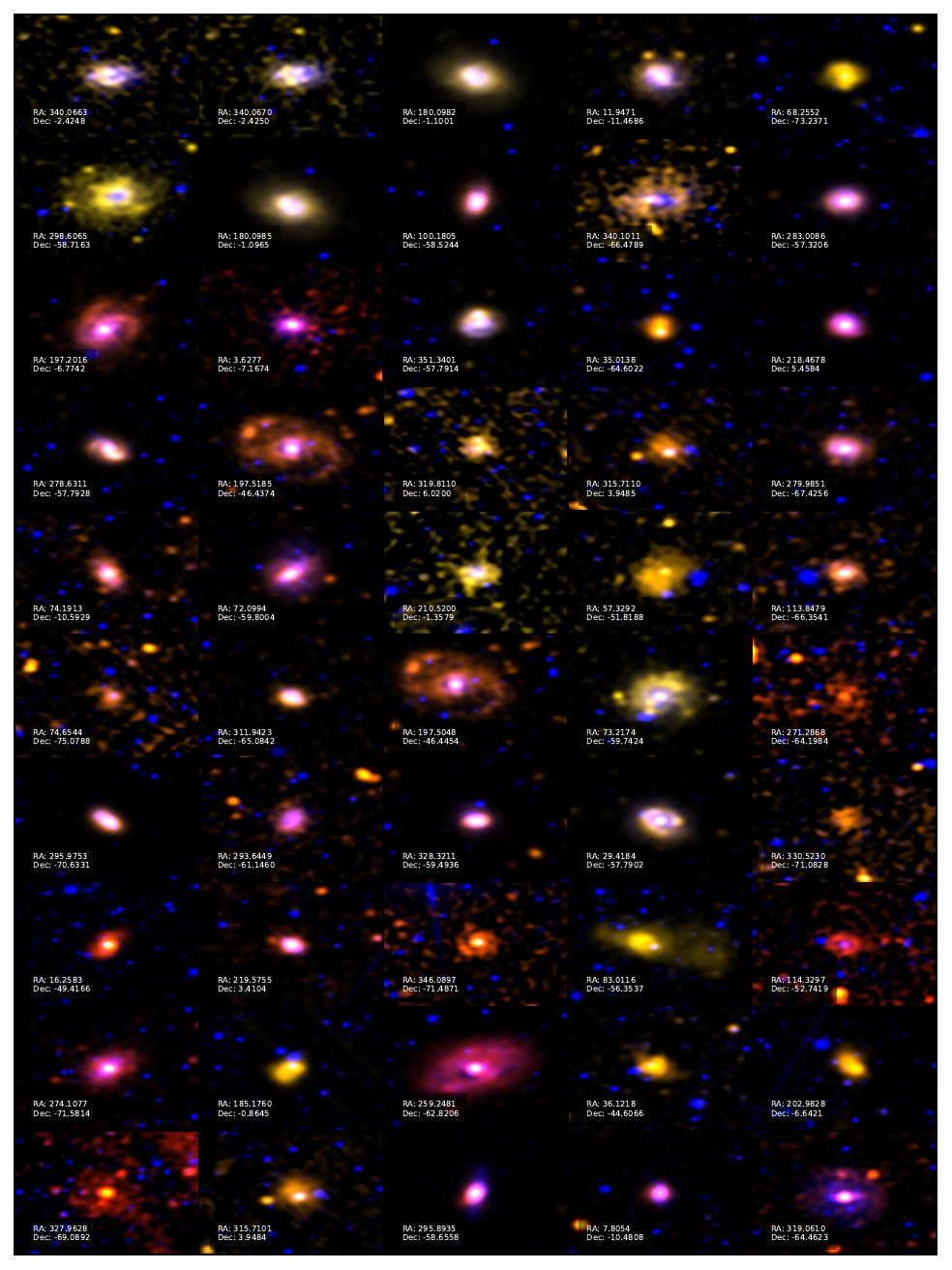}
\caption{Top-50 EMUSE output for the text query, ``Resolved star forming radio galaxy". Positions in Table~\ref{TAB:starforming} are used here for $5^{\prime}\times5^{\prime}$ cutout images with radio-radio-infrared channels.} 
\label{FIG:starforming}
\end{figure*}

\begin{table*}[!htbp]
    \centering
    \begin{tabular}{c c c c c c}
    \hline
    \textbf{SBID} & \textbf{RA (Degrees)} & \textbf{Dec (Degrees)} & \textbf{Integrated Flux (mJy)} & \textbf{CatWISE Potential Host} & \textbf{Probability} \\
    \hline
    50786 & 316.61292 & -4.78572 & 10.53 & J210627.10-044708.6 & 1.00 \\
    53293 & 18.19064 & -32.22042 & 5.78 & J011245.75-321313.5 & 1.00 \\
    52219 & 202.43153 & -4.45410 & 45.05 & J132943.56-042714.7 & 1.00 \\
    50413 & 214.07986 & -30.43106 & 89.79 & J141619.16-302551.8 & 1.00 \\
    46959 & 248.13356 & -67.51922 & 15.87 & nan & 1.00 \\
    51956 & 215.66485 & -32.45638 & 16.27 & J142239.56-322722.9 & 0.99 \\
    46943 & 217.75088 & -27.24095 & 8.59 & J143100.21-271427.4 & 0.99 \\
    52145 & 110.91825 & -56.29562 & 6.02 & J072340.37-561744.2 & 0.99 \\
    54105 & 262.17056 & -58.22261 & 33.48 & J172840.93-581321.3 & 0.99 \\
    51948 & 202.43108 & -45.65679 & 39.95 & J132943.45-453924.4 & 0.99 \\
    50049 & 128.88849 & -0.12402 & 76.55 & J083533.23-000726.4 & 0.99 \\
    46984 & 226.25434 & -29.33961 & 92.99 & J150501.04-292022.5 & 0.99 \\
    53293 & 16.15859 & -32.80721 & 19.25 & J010438.06-324825.9 & 0.99 \\
    51845 & 135.19085 & -65.74613 & 24.90 & J090045.80-654446.0 & 0.99 \\
    51430 & 316.38351 & -8.14548 & 266.17 & J210532.04-080843.7 & 0.99 \\
    51948 & 200.09450 & -48.31177 & 26.21 & J132022.68-481842.3 & 0.99 \\
    54769 & 278.69927 & -70.52948 & 16.38 & J183447.82-703146.1 & 0.99 \\
    51853 & 90.49432 & -61.78667 & 19.35 & J060158.63-614712.0 & 0.99 \\
    54926 & 260.52841 & -65.20608 & 30.05 & nan & 0.99 \\
    51959 & 107.88512 & -59.62079 & 56.10 & J071132.42-593714.8 & 0.99 \\
    46966 & 25.48851 & -47.38656 & 135.70 & J014157.24-472311.6 & 0.99 \\
    59862 & 338.72276 & -52.02211 & 17.26 & J223453.46-520119.5 & 0.99 \\
    51559 & 196.72903 & -6.66626 & 9.97 & J130654.96-063958.5 & 0.99 \\
    53566 & 334.88363 & -70.04507 & 99.03 & J221932.07-700242.2 & 0.99 \\
    45761 & 331.92321 & -58.53195 & 13.55 & J220741.57-583155.0 & 0.99 \\
    53183 & 31.39560 & -3.51602 & 14.33 & J020534.94-033057.6 & 0.99 \\
    46984 & 225.38555 & -26.05533 & 23.63 & J150132.53-260319.1 & 0.99 \\
    59253 & 32.19052 & -9.28769 & 8.62 & J020845.72-091715.6 & 0.99 \\
    54944 & 317.80591 & -58.32017 & 17.78 & J211113.41-581912.5 & 0.99 \\
    46978 & 89.13536 & -72.11116 & 145.02 & nan & 0.99 \\
    51927 & 103.99102 & -55.89982 & 86.91 & J065557.84-555359.3 & 0.99 \\
    54099 & 251.46808 & -70.70401 & 4.41 & J164552.33-704214.4 & 0.99 \\
    50182 & 207.68442 & -1.60279 & 66.31 & J135044.26-013610.0 & 0.99 \\
    51434 & 319.58953 & 1.68626 & 51.62 & nan & 0.99 \\
    53304 & 106.05205 & -71.92890 & 21.78 & J070412.49-715544.0 & 0.99 \\
    52179 & 309.15719 & -19.99941 & 32.53 & J203637.72-195957.8 & 0.99 \\
    53557 & 347.29875 & -69.63885 & 21.84 & nan & 0.99 \\
    55326 & 40.19354 & -4.85017 & 97.93 & J024046.45-045100.6 & 0.99 \\
    59159 & 50.59763 & -44.95993 & 12.03 & J032223.43-445735.7 & 0.99 \\
    55326 & 38.39934 & -6.26939 & 9.57 & J023335.84-061609.8 & 0.99 \\
    53313 & 3.18376 & -9.52691 & 24.55 & J001244.10-093136.8 & 0.98 \\
    47136 & 27.32598 & -49.39718 & 16.81 & J014918.23-492349.8 & 0.98 \\
    54926 & 270.53502 & -63.61610 & 20.11 & J180208.40-633657.9 & 0.98 \\
    54802 & 301.48566 & -68.00271 & 9.24 & nan & 0.98 \\
    51797 & 322.61649 & -54.85862 & 380.63 & J213027.95-545131.0 & 0.98 \\
    54104 & 123.48123 & -56.99331 & 71.76 & J081355.49-565935.9 & 0.98 \\
    46957 & 62.52834 & -70.44637 & 26.79 & J041006.80-702646.9 & 0.98 \\
    46980 & 261.24429 & -66.32601 & 59.09 & nan & 0.98 \\
    51448 & 319.58879 & 1.68616 & 51.57 & nan & 0.98 \\
    51430 & 317.08857 & -8.84341 & 38.37 & J210821.25-085036.2 & 0.98 \\
    46925 & 31.93579 & -66.12144 & 19.10 & J020744.58-660717.1 & 0.98 \\
    \hline
    \end{tabular}
    \caption{Top-50 EMUSE output for image query shown on the left panel of Figure~\ref{FIG:imagequeries}.}
    \label{TAB:fr2}
\end{table*}

\begin{figure*}[!htbp]
\centering
\includegraphics[width=0.98\textwidth]{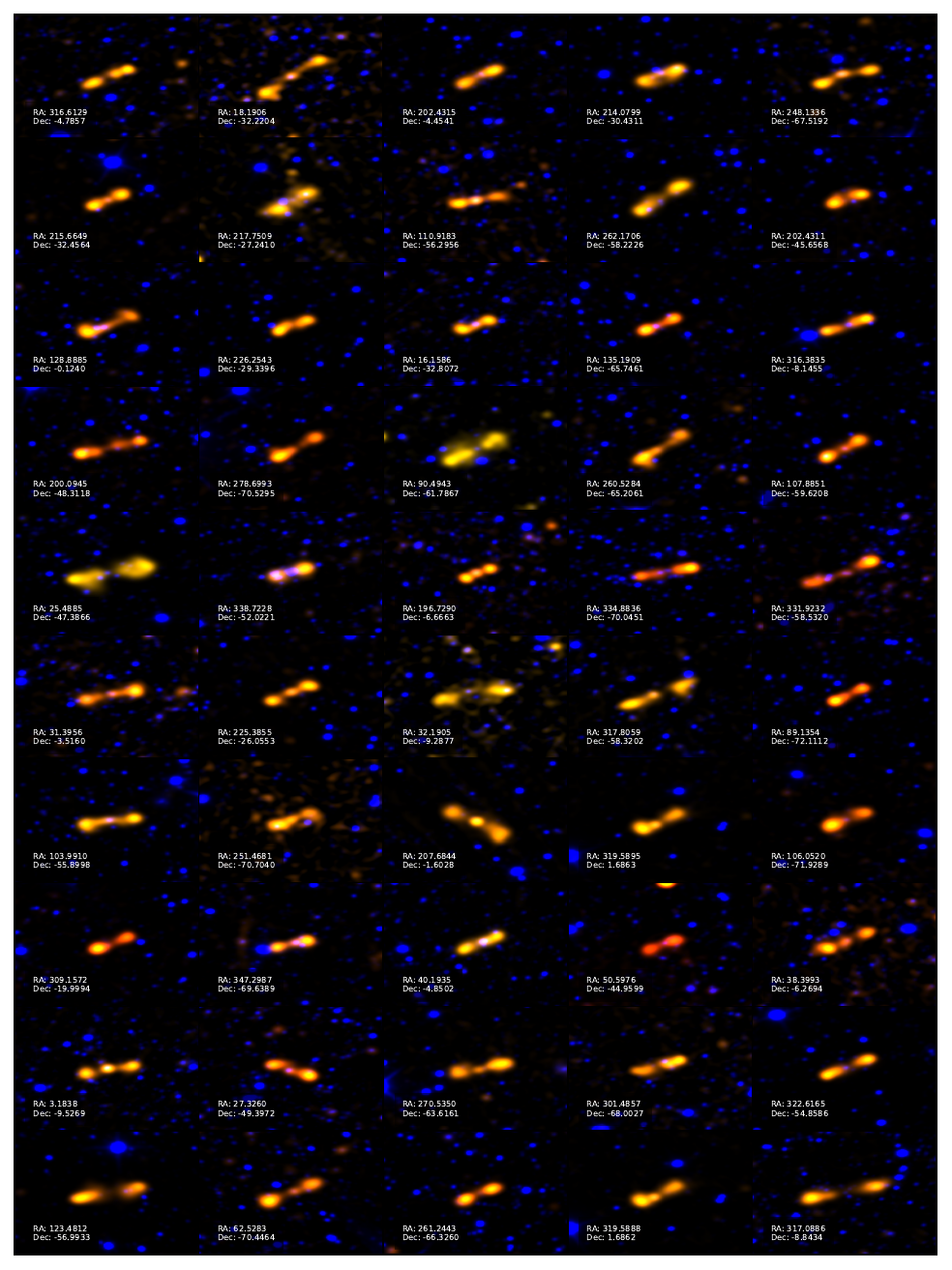}
\caption{Top-50 EMUSE output for image query shown on the left panel of Figure~\ref{FIG:imagequeries}. Positions in Table~\ref{TAB:fr2} are used here for $5^{\prime}\times5^{\prime}$ cutout images with radio-radio-infrared channels.} 
\label{FIG:fr2}
\end{figure*}

\begin{table*}[!htbp]
    \centering
    \begin{tabular}{c c c c c c}
    \hline
    \textbf{SBID} & \textbf{RA (Degrees)} & \textbf{Dec (Degrees)} & \textbf{Integrated Flux (mJy)} & \textbf{CatWISE Potential Host} & \textbf{Probability} \\
    \hline
    46984 & 227.02211 & -25.77369 & 7.61 & J150805.30-254625.2 & 1.00 \\
    50538 & 77.63662 & -58.42155 & 2.17 & J051032.78-582517.5 & 1.00 \\
    51956 & 211.64795 & -34.31162 & 9.76 & J140635.50-341841.8 & 0.99 \\
    51962 & 211.78771 & -9.28361 & 4.36 & J140709.04-091700.9 & 0.97 \\
    59094 & 49.29686 & -52.15329 & 1.51 & J031711.24-520911.8 & 0.97 \\
    46959 & 241.29206 & -70.51253 & 0.62 & J160510.09-703045.0 & 0.96 \\
    54773 & 11.76460 & -11.87260 & 42.27 & J004703.50-115221.3 & 0.96 \\
    50048 & 20.79271 & -54.33171 & 0.95 & nan & 0.95 \\
    53304 & 105.78548 & -70.59927 & 1.27 & J070308.51-703557.3 & 0.94 \\
    46978 & 80.58389 & -71.56971 & 11.67 & J052220.13-713410.9 & 0.94 \\
    46955 & 266.63387 & -68.26133 & 1.94 & J174632.12-681540.7 & 0.94 \\
    50048 & 20.81822 & -54.34069 & 3.21 & nan & 0.94 \\
    51434 & 318.06099 & -0.39710 & 1.69 & nan & 0.94 \\
    54926 & 269.71410 & -64.15688 & 0.87 & J175851.38-640924.7 & 0.94 \\
    54098 & 118.77421 & -49.65293 & 0.49 & J075505.81-493910.5 & 0.93 \\
    45781 & 334.25553 & -57.67611 & 3.50 & nan & 0.93 \\
    51430 & 316.07299 & -8.07676 & 41.06 & J210417.51-080436.3 & 0.93 \\
    59804 & 137.82948 & -17.94543 & 14.55 & J091119.07-175643.5 & 0.93 \\
    46951 & 16.90893 & -51.46922 & 1.83 & J010738.14-512809.1 & 0.92 \\
    45761 & 327.34774 & -59.37191 & 4.03 & nan & 0.92 \\
    53210 & 99.31469 & -48.46778 & 1.65 & J063715.52-482804.0 & 0.92 \\
    59560 & 175.97084 & -1.69125 & 7.73 & J114353.00-014128.5 & 0.92 \\
    53313 & 5.20887 & -9.26766 & 1.65 & nan & 0.92 \\
    54103 & 307.24684 & -69.51609 & 0.89 & J202859.24-693057.9 & 0.92 \\
    59253 & 32.81166 & -7.86003 & 1.10 & J021114.79-075136.0 & 0.92 \\
    53314 & 109.11890 & -66.81888 & 1.75 & nan & 0.92 \\
    51574 & 281.18582 & -57.64771 & 1.63 & J184444.59-573851.7 & 0.91 \\
    52219 & 204.89026 & -6.51725 & 0.64 & J133933.66-063102.0 & 0.91 \\
    54098 & 117.49202 & -51.49752 & 1.17 & J074958.08-512951.0 & 0.91 \\
    46978 & 83.45844 & -72.04856 & 185.75 & J053350.02-720254.8 & 0.91 \\
    50427 & 245.71940 & -64.27653 & 4.75 & J162252.65-641635.5 & 0.91 \\
    54103 & 317.34480 & -68.78581 & 1.43 & nan & 0.91 \\
    51931 & 82.57955 & -56.87337 & 0.98 & J053019.09-565224.1 & 0.91 \\
    51932 & 200.62191 & -6.58950 & 3.34 & J132229.25-063522.1 & 0.91 \\
    46951 & 15.08030 & -53.06277 & 0.54 & J010019.27-530345.9 & 0.91 \\
    50413 & 210.26688 & -30.32623 & 47.21 & J140104.05-301934.4 & 0.91 \\
    59253 & 31.31543 & -11.52721 & 1.95 & J020515.70-113137.9 & 0.90 \\
    51403 & 128.64065 & -62.13250 & 2.77 & J083433.75-620757.0 & 0.90 \\
    59612 & 327.61388 & -69.70213 & 0.97 & J215027.33-694207.6 & 0.90 \\
    59159 & 49.09901 & -48.45771 & 2.08 & nan & 0.90 \\
    50180 & 12.24043 & -47.22998 & 6.56 & nan & 0.90 \\
    51403 & 127.05199 & -59.82145 & 2.69 & nan & 0.90 \\
    54926 & 271.28683 & -64.19840 & 0.78 & J180508.83-641154.2 & 0.90 \\
    51434 & 318.06051 & -0.39312 & 0.84 & J211214.52-002335.2 & 0.90 \\
    50049 & 128.60151 & 0.08055 & 3.78 & J083424.36+000449.9 & 0.89 \\
    50413 & 215.14208 & -29.25002 & 5.52 & nan & 0.89 \\
    45781 & 334.25606 & -57.67157 & 1.04 & nan & 0.89 \\
    51797 & 319.29061 & -55.94363 & 8.14 & J211709.74-555637.0 & 0.89 \\
    52121 & 314.22566 & -64.96138 & 4.06 & J205654.15-645740.9 & 0.89 \\
    54099 & 257.05483 & -71.22234 & 4.17 & J170813.16-711320.4 & 0.89 \\
    46957 & 59.50084 & -73.62135 & 2.93 & nan & 0.89 \\
    \hline
    \end{tabular}
    \caption{Top-50 EMUSE output for image query shown on the right panel of Figure~\ref{FIG:imagequeries}.}
    \label{TAB:orc1}
\end{table*}

\begin{figure*}[!htbp]
\centering
\includegraphics[width=0.98\textwidth]{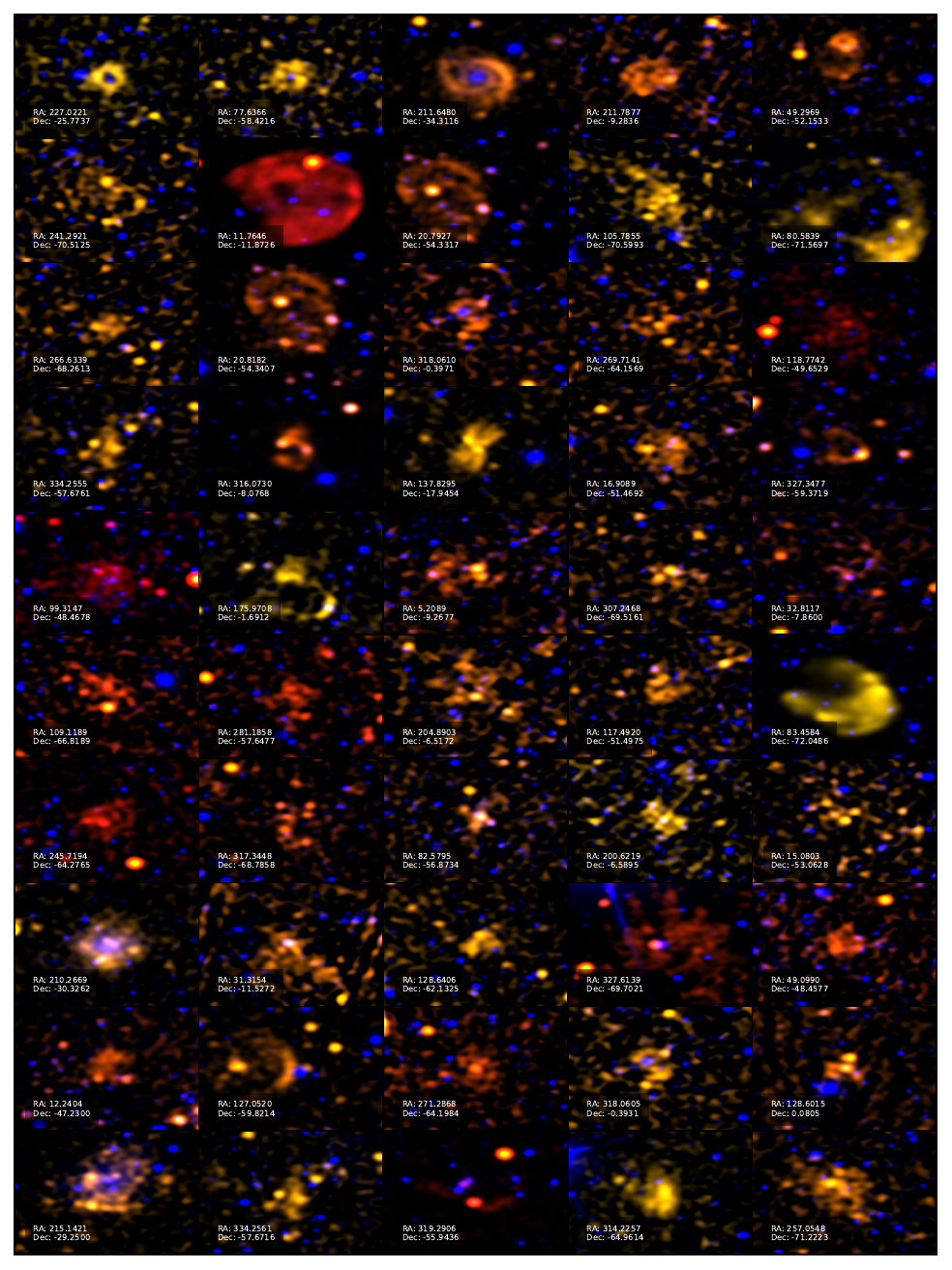}
\caption{Top-50 EMUSE output for image query shown on the right panel of Figure~\ref{FIG:imagequeries}. Positions in Table~\ref{TAB:orc1} are used here for $5^{\prime}\times5^{\prime}$ cutout images with radio-radio-infrared channels.} 
\label{FIG:orc1}
\end{figure*}

\begin{table*}[htbp]
\centering
\caption{Examples of the expanded text descriptions for the main radio source classes. These, along with similar variations based on subcategories and special features, are used to fine-tune the OpenCLIP model.}
\begin{tabularx}{\textwidth}{@{}lX@{}}
\toprule
\textbf{Main Category} & \textbf{Expanded Text Descriptions} \\
\midrule
FR-I & An image of FR-I; fr1; double lobed FR-I radio galaxy; Fanaroff-Riley type I radio galaxy; FR-I radio galaxies are characterized by edge-dimmed radio morphology, where the brightest emission is near the core and gradually fades outward along the jets. \\
\addlinespace
FR-II & An image of FR-II; double-lobed FR-II; double-lobed FR-II type radio galaxy; double-lobed Fanaroff-Riley type II radio galaxy; FR-II radio galaxies exhibit edge-brightened radio morphology, with the brightest emission located at the outer edges of the lobes, often forming distinct hotspots. \\
\addlinespace
FR-x & An image of FR-X; frx; FRx radio galaxy that has morphology in-between FR-I and FR-II types but can't be determined due to lack of telescope sensitivity and resolution; Fanaroff-Riley type X, an intermediate type between type I and type II radio galaxies with morphological structure in-between FR-I and FR-II types. \\
\addlinespace
R & An image of Single-Peak R/DJS; djs; double jet radio source; resolved DJS or R radio galaxy with a single peak visible in the centre. \\
\addlinespace
Peculiar & An image of Complex/Peculiar; ORC; ORC that is an Odd Radio Circle; peculiar radio galaxy classified as an Odd Radio Circle; ring-like structure seen in radio with no corresponding emission in other wavelengths. \\
\bottomrule
\end{tabularx}
\label{TAB:expanded_text}
\end{table*}

\label{lastpage}
\end{document}